\documentclass[twocolumn,showpacs,letterpaper,preprintnumbers,amsmath,amssymb]{revtex4}
\usepackage{psfrag}
\usepackage{epsfig}
\usepackage{graphicx}
\begin{document}
\title{Velocity difference statistics in turbulence}
\author{Sunghwan Jung}
\email{sunnyjsh@chaos.utexas.edu}
\author{Harry L. Swinney}
\affiliation{Center for Nonlinear Dynamics and Department of Physics, \\
 The University of Texas at Austin, Austin, Texas 78712 USA.}
\date{\today}

\begin{abstract}
We unify two approaches that have been taken to explain the
non-Gaussian probability distribution functions (PDFs) obtained in
measurements of longitudinal velocity differences in turbulence, and
we apply our approach to Couette-Taylor turbulence data.  The first
approach we consider was developed by Castaing and coworkers, who
obtained the non-Gaussian velocity difference PDF from a
superposition of Gaussian distributions for subsystems that have a
particular energy dissipation rate at a fixed length scale [Castaing
et al., {\it Physica D} {\bf 46}, 177 (1990)]. Another approach was
proposed by Beck and Cohen, who showed that the observed PDFs can be
obtained from a superposition of Gaussian velocity difference PDFs
in subsystems conditioned on the value of an intensive variable
(inverse ``effective temperature'') in each subsystem [Beck and
Cohen, {\it Physica A} {\bf 322}, 267 (2003)]. The intensive
variable was defined for subsystems assuming local thermodynamic
equilibrium, but no method was proposed for determining the size of
a subsystem. We show that the Castaing and Beck-Cohen methods are
related, and we present a way to determine subsystem size in the
Beck-Cohen method. The application of our approach to Couette-Taylor
turbulence (Reynolds number $540~000$) yields a log-normal
distribution of the intensive parameter, and the resultant velocity
difference PDF agrees well the observed non-Gaussian velocity
difference PDFs.

\end{abstract}



\maketitle

\section{Introduction}

In Kolmogorov's 1941 theory (K41), the energy in fully developed
three-dimensional turbulence cascades from large scales to small
scales where it is dissipated \cite{kolmogorov41}.  Turbulence in
the cascade (the inertial range) is characterized by the probability
distribution function (PDF) $P(\delta v_r)$ for longitudinal
velocity differences over a distance $r$, $\delta v_r(x) = \hat{e}_r
\cdot [\vec{v}({x} +{r}) - \vec{v}({x}) ]$), where $\hat{e}_r$ is
the direction of separation \cite{frisch}. For $r$ approaching the
integral scale where energy is injected, the PDF is Gaussian, while
in the inertial range extending down to the dissipation scale
$\eta$, intermittent large fluctuations lead to a non-Gaussian PDF
with approximately exponential tails \cite{LN_van70}.

Kolmogorov assumed a constant energy dissipation rate per unit
volume, $\varepsilon$ \cite{kolmogorov41}. In 1944 Landau
\cite{landau} suggested that fluctuations of $\varepsilon$ averaged
at scale $r$, $\varepsilon_r(\vec x,t) \left( = \int_{\vec x}^{\vec
x + \vec r } \varepsilon(\vec x',t) d \vec x' \right)$ play a key
role in turbulence. Such fluctuations were subsequently observed in
many experiments~ \cite{KO_batchelor_49,KO_grant_62,KO_meneveau_91}.
In 1962 Kolmogorov~\cite{kolmogorov62} and Obukhov~\cite{obukhov62}
proposed a log-normal distribution of $\varepsilon_r$ in the
inertial range. The log-normal distribution was obtained in
subsequent experiments and numerical simulations
$\varepsilon_r$~\cite{KO158,KO30,KO123,KO27}. The non-Gaussian PDF
of $\delta v_r$ and the log-normal PDF of $\varepsilon_r$
characterize turbulent flows.

Different approaches have been taken by Castaing et
al.~\cite{castaing90} and by Beck and Cohen~\cite{beck03} to
understand the non-Gaussian $P(\delta v_r)$. Castaing et al.
assumed that subsystems have different values of $\varepsilon_r$,
but the subsystems have Gaussian PDFs of $\delta v_r$; this
assumption is supported by experiments.

Beck and Cohen took a statistical mechanics approach, assuming that
subsystems have a well-defined ``effective temperature'', which for
turbulent flow is identified with the variance of $\delta v_r$. The
resultant $P(\delta v_r)$ depends on the statistics of the
distribution for the inverse effective temperatures in the
subsystems. This dependence of the statistical distribution
$P(\delta v_r)$ on the statistical distribution of subsystems led
Beck and Cohen to call their approach {\it
superstatistics}~\cite{beck03}.

In this paper we note that the approaches of Castaing et al. and
Beck and Cohen are both based on Bayes' theorem,
\begin{equation}
P(x) = \int P(x|y) P(y) dy,
\end{equation}
which is used to obtain the non-Gaussian $P(\delta v_r)$ from a
conditional mixing of Gaussian PDFs in subsystems. However, the
subsystems are chosen differently in the two approaches.

We propose a method that does not require a determination of
$\varepsilon_r$ from experimental data, nor does it require a
fitting parameter to obtain the effective temperature PDF. We show
that subsystems with Gaussian statistics can be chosen by examining
moments of velocity difference distributions in the subsystems. Our
method, which involves no fitting parameters, leads to predictions
for the non-Gaussian $P(\delta v_r)$ that are in accord with data
for turbulent Couette-Taylor flow~\cite{lewis99}.

In Section \ref{sec:sup_super} we present the Castaing et al. and
Beck and Cohen methods, and in Section \ref{sec:sup_exp} we
describe the Couette-Taylor experiments and present results for
$P(\delta v_r)$. Section \ref{sec:sup_results} shows how
subsystems can be systematically chosen to obtain a prediction for
$P(\delta v_r)$. The conclusions are presented in Section
\ref{sec:sup_conclusion}.


\section{Theory} \label{sec:sup_super}

\subsection{Method of Castaing et al.}

Castaing et al.~\cite{castaing90} started with the observation from
their experiments that that velocity difference distributions for a
given $\varepsilon_r$ are Gaussian, and that $\varepsilon_r$ is
described by a log-normal
distribution~\cite{castaing90,castaing93,LN90,LN175,LN161,LN146}.
The log-normal distribution for $\varepsilon_r$ has also been
obtained for $\varepsilon_r$ in other experiments on fully developed
turbulence~\cite{LN112,LN65,LN90,LN93}, and in analyses of images of
cloud patterns \cite{LN68}, effective temperature fields in
turbulence \cite{LN80}, and magnetic fields in solar winds
\cite{LN69}.

To describe the evolution of $P(\delta v_r)$ from Gaussian at large
scales to non-Gaussian at small scales \cite{LN211,LN135,LN128},
Castaing et al. proposed \cite{castaing90,castaing93}
\begin{equation} \label{eq:sup_kolcas}
    P ( {\delta v_r} ) =  \int  P
    ( \varepsilon_r) P(\delta v_r | \varepsilon_r) d
    \varepsilon_r .
\end{equation}
The conditional PDF $P(\delta v_r | \varepsilon_r)$ in Eq.
(\ref{eq:sup_kolcas}) is assumed to be a Gaussian distribution,
$P(\delta v_r | \varepsilon_r) = e^{-(\delta v_r)^2/(r
\varepsilon_r)^{2/3}}$, in accord with experimental observations
\cite{KO50,LN183,LN_wait92}. Kolmogorov \cite{kolmogorov62}, Obukhov
\cite{obukhov62} and Castaing \cite{castaing90} assumed a log-normal
distribution of $\varepsilon_r$,
\begin{equation} \label{eq:sup_intro_4}
P ( \varepsilon_r) = \frac{1}{\lambda_{\varepsilon} (2 \pi)^{1/2}
\varepsilon_r} \exp \left(- \frac{(\ln \varepsilon_r
-m_{\varepsilon})^2}{2 \lambda_{\varepsilon}^2} \right),
\end{equation}
where $m_{\varepsilon}$ and $\lambda_{\varepsilon}$ are respectively
the mean and the standard deviation of $\ln \varepsilon_r$.

A difficulty in applying the approach of Castaing et al. is that
energy dissipation rate at length scale $r$, $\varepsilon_r$, is not
directly measured in experiments. By assuming homogeneous and
isotropic conditions, $\varepsilon_r(x)$ is defined as $15 \nu
\int_x ^{x+r} \left( \partial v / \partial x \right)^2 dx$. In
practice, $\varepsilon_r$ is determined from time series data,
\begin{equation} \label{eq:sup_surro}
\varepsilon_r = \frac{15 \nu}{(\Delta x)^2} \sum_{i=1}^{N-1}
[v(x_{i+1}) -v(x_i)]^2,
\end{equation}
\noindent where $\Delta x (\equiv x_{2} - x_1)$ is the sampling
separation the summation $i$ is over subsystems and $x_N -x_1 =r$
\cite{LN_wait92,LN_wait96,LN_wait70}. Even with this assumption,
determination of $\varepsilon_r(x)$ is difficult because of errors
in evaluating the derivative from velocity data. Further error
arises from the application of the Taylor frozen hypothesis at high
frequencies~ \cite{LN_fisher64,LN43,LN93,KO140,
KO_wyngaard_70,LN_pinton94,
LN_wyngaard77,LN_antonia80,LN_lumley65,LN_champagne78}.

\subsection{Superstatistics of Beck and Cohen}

Beck and Cohen's statistical approach considers a system far from
thermodynamic equilibrium to consist of subsystems in local
thermodynamic equilibrium \cite{beck03}. Each subsystem has a
well-defined ``effective temperature'', but the subsystem effective
temperatures need not be the same since the whole system is not in
equilibrium. Beck and Cohen identify $(\delta v_r)^2$ with the
kinetic energy of eddies of size $r$, $E(\delta v_r) = \frac{1}{2}
(\delta v_r)^2$, and the variance of $\delta v_r$ is identified with
an inverse effective temperature $\beta$~\cite{LN155}, given for a
subsystem of size $d$ by

\begin{equation}
\beta_d = \frac{1}{ \langle (\delta v_r)^2 \rangle_d
- (\langle \delta v_r \rangle_d)^2}, \label{eq:beta}
\end{equation}
where $\langle \cdot \rangle_d$ is an average over the size $d$.
Then we have
\begin{equation}
\label{eq:super_tem}
    P({\delta v_r} ) = \int^{\infty}_0 P(\beta_d)
    P (\delta v_r| \beta_d) d \beta_d.
\end{equation}
where $P(\beta_d)$ is the distribution of inverse effective
temperature in subsystems of size $d$.

A particular choice of $P(\beta_d)$, the $\chi^2$ distribution,
leads to the distribution associated with the nonextensive
statistical mechanics of Tsallis, $P(E) = (1+\beta (q-1)
E)^{-1/(q-1)}$, where $q$ is a parameter characterizing the
nonextensivity [$S(1+2) = S(1)+S(2) + (1-q) S(1) \cdot S(2)$, where
$S$ is entropy function.] ~\cite{tsallis88,beck03}. A phenomenology
similar to Beck and Cohen's was used in earlier oceanographic
analysis that described the global non-Gaussian distribution of
ocean surface velocity as a mixture of local Gaussians with
$\chi^2$-distributed variance \cite{LN58,LN77}. The method of Beck
and Cohen has been applied to fully developed turbulence
\cite{LN02,beck04} by introducing a fitting parameter to determine
the PDF of inverse effective temperature, rather than by directly
measuring the PDF of inverse effective temperature.

The Beck-Cohen method requires that the size $d$ should be large
compared to the distance $r$ separating two points, and $d$ should
also be large enough so the subsystems contain enough data points to
yield good statistics, but $d$ must also be small enough so that
subsystems are each described by a Gaussian distribution.  Beck
determined the size of $d$ using a fitting parameter involving the
kurtosis of $P({\delta v_r} )$~\cite{beck04}.

\subsection{Unified view of PDFs}

The Castaing and Beck-Cohen methods are similar except in the way
they divide a system into subsystems. Castaing et al. sample
velocity differences conditioned by the averaged energy dissipation
rate $\varepsilon_r$, while Beck and Cohen use velocity differences
conditioned by the inverse effective temperature $\beta_d$. Castaing
et al. need one fixed length scale, the separation distance $r$
between two points; $\delta v_r$ and $\varepsilon_r$ are defined at
this scale and are related through Bayes' theorem.  The Beck-Cohen
method involves two length scales, the distance $r$ separating two
points and the size $d$ of the subsystems in the statistical
analysis.

The Castaing and Beck-Cohen methods can be connected if the two
conditioning variables ($\varepsilon_r$ and $\beta_d$) are
correlated. Using Eq. (\ref{eq:sup_kolcas}) and Bayes' theorem,
we convert Castaing's method into Beck-Cohen's method,
\begin{eqnarray}
&&P ( {\delta v_r} )= \int_0^{\infty}  P(\delta v_r | \varepsilon_r)
P( \varepsilon_r)  d \varepsilon_r \nonumber \\
&=& \int_0^{\infty} \int_0^{\infty} P(\delta v_r | \beta_d)
P(\beta_d| \varepsilon_r) d
\beta_d P( \varepsilon_r)  d \varepsilon_r \nonumber \\
&=&  \int_0^{\infty}  P(\delta v_r | \beta_d) \left[ \int_0^{\infty}
P(\beta_d| \varepsilon_r)  P( \varepsilon_r) d \varepsilon_r \right]
d \beta_d \label{eq:sup_uni_1} \\
&=&  \int_0^{\infty}  P(\delta v_r | \beta_d)  P( \beta_d)  d
\beta_d . \label{eq:sup_uni_2}
\end{eqnarray}
Now, let's assume a log-normal distribution of $\beta_d$ at
the fixed $\varepsilon_r$,
\begin{equation} \label{eq:sup_uni_3}
P(\beta_d | \varepsilon_r) \propto \frac{1}{ \beta_d} \exp \left[ -
\frac{ ( \ln \beta_d - a \ln \varepsilon_r)^2 }{ 2 {\lambda_t}^2}
\right],
\end{equation}
where $\lambda_t$ is the standard deviation of $\ln \beta_d$
conditioned to $\varepsilon_r$, and $a$ is a parameter. Using Eqs.
(\ref{eq:sup_intro_4}), (\ref{eq:sup_uni_1}), and
(\ref{eq:sup_uni_3}), we have
\begin{eqnarray}
& P ({\delta v_r})& \propto \int_0^{\infty} P(\delta v_r| \beta_d)
 \int_0^{\infty} \exp \left( - \frac{(\ln \beta_d  - a \ln
\varepsilon_r)^2}{2 {\lambda_t}^2} \right) \nonumber \\
&& \times \exp  \left( - \frac{(\ln \varepsilon_r
-m_{\varepsilon})^2}{2 \lambda_{\varepsilon}^2} \right) d (\ln
\varepsilon_r) d (\ln \beta_d) \nonumber \\
&\propto& \int_0^{\infty} P(\delta v_r| \beta_d) \exp \left(
-\frac{(\ln \beta_d -m)^2}{ \lambda_{\varepsilon}^2 \lambda_{t}^2}
\right) d (\ln \beta_d). \nonumber \\
\end{eqnarray}
Thus with the assumption of a log-normal distribution of $\beta_d$
conditioned on $\varepsilon_r$, we have that Castaing's method is
equivalent to Beck-Cohen's method. In Section \ref{sec:sup_comp},
the log-normal PDF of $P(\beta_d | \varepsilon_r)$ is verified in
experiments.


\begin{figure}
\begin{center}
  \psfrag{cm}{$x$ (cm)}
  \psfrag{Y}{$\delta v_{r}$ (cm/s)}
  \includegraphics[width=\linewidth]{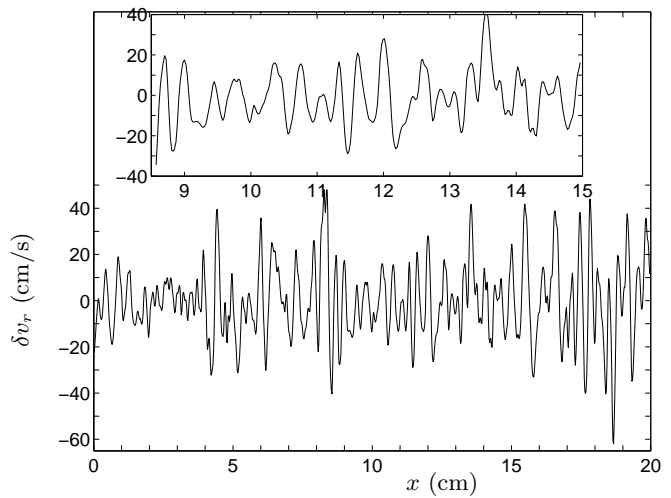}
\end{center}
\caption{ An example of the Couette-Taylor velocity difference data,
 obtained by subtracting velocities at two points with a
 separation $r = 46 \eta = 0.134$ cm, where $\eta$ is the Kolmogorov
 length scale. The inset shows the velocity differences on a finer
 length scale. }
 \label{fig:window}
\end{figure}

\section{Experiment} \label{sec:sup_exp}
We describe here an experiment on turbulent Couette-Taylor flow by
Lewis and Swinney~\cite{lewis99,lathrop92}, and in the next section
we will analyze data from this experiment to deduce $P(\beta)$ and a
prediction for $P({\delta v_r})$. The fluid was contained in the
annular region between two concentric cylinders with an inner radius
of $b = 22.085 \mathrm{~cm}$ and an outer radius of $a = 15.999
\mathrm{~cm}$; thus the ratio of inner to outer radius was 0.724.
The height of the annulus was 69.5 cm, which yields a value of 11.4
for the ratio of height to the gap. The inner cylinder angular
rotation rate $\Omega$ was $8 \times 2 \pi$ rad/s; the outer
cylinder was at rest. The ends of the annulus rotated at the same
rate as the inner cylinder. The fluid was water with a viscosity
$\nu$ of $0.00968 \mathrm {~cm^2/s}$ at the working effective
temperature. Defining the Reynolds number as $Re = {\Omega a (b-a)}/
{ \nu}$ yields for the Reynolds number $540~000$ \cite{lewis99}.

A hot film probe was used to measure the time dependence of the
azimuthal component of the velocity in the center of the gap at a
distance $4.35 \mathrm{~cm}$ above mid-height of the annulus. The
Taylor frozen turbulence hypothesis was used to convert the
velocity time series data to velocity field data. The turbulent
intensity (the ratio of the root mean squared velocity to the mean
velocity) was less than 6\%.

The uncertainties shown on our graphs correspond to the standard
deviation of 20 independent experiments. The velocity measurements
were made with a sampling rate 2500 times the inner cylinder
rotation frequency; this corresponds to a spatial separation of
0.017 cm between successive velocity values. The longitudinal
velocity differences $\delta v_r$ that we analyze are for points
separated by a small distance, $r = 0.134$ cm, where the probability
distribution function has approximately exponential tails
~\cite{lewis99}. An example of the measurements of $\delta v_r(t)$
is shown in Fig. \ref{fig:window}. The separation $r = 0.134$ cm
corresponds to $46 \eta$, where $\eta$ is Kolmogorov
scale~\cite{lewis99}. (The Kolmogorov dissipation scale was obtained
by calculating the dissipation from energy spectra: $\eta \equiv
(\nu/\varepsilon)^{1/4}$, where the dissipation rate is given by
$\varepsilon = 15 \nu \int k^2 E(k) dk $ \cite{lewis99}.)  The
window size $d$ we use for determining the local inverse effective
temperature $\beta$ is typically 0.9 cm, nearly an order of
magnitude larger than the value of $r$.


\section{Results} \label{sec:sup_results}

\subsection{Probability density function of inverse effective temperature} \label{sec:sup_distributions}

\begin{figure*}
 \centering
 \begin{center}
   \psfrag{Pbeta}{$P(\beta_d)$}
   \psfrag{b1}{$\beta_{\mathrm{0.9~cm}}/(\beta_{\mathrm{0.9~cm}})_{rms}$ }
   \psfrag{b15}{$\beta_{\mathrm{3~cm}}/(\beta_{\mathrm{3~cm}})_{rms}$}
   \psfrag{a}{(a)} \psfrag{chi}{$\chi^2$}
   \psfrag{b}{(b)} \psfrag{ln}{log-normal}
   \psfrag{Eb}{Difference}
   \psfrag{c}{(c)}\psfrag{d}{(d)}
  \includegraphics[width=\linewidth]
     {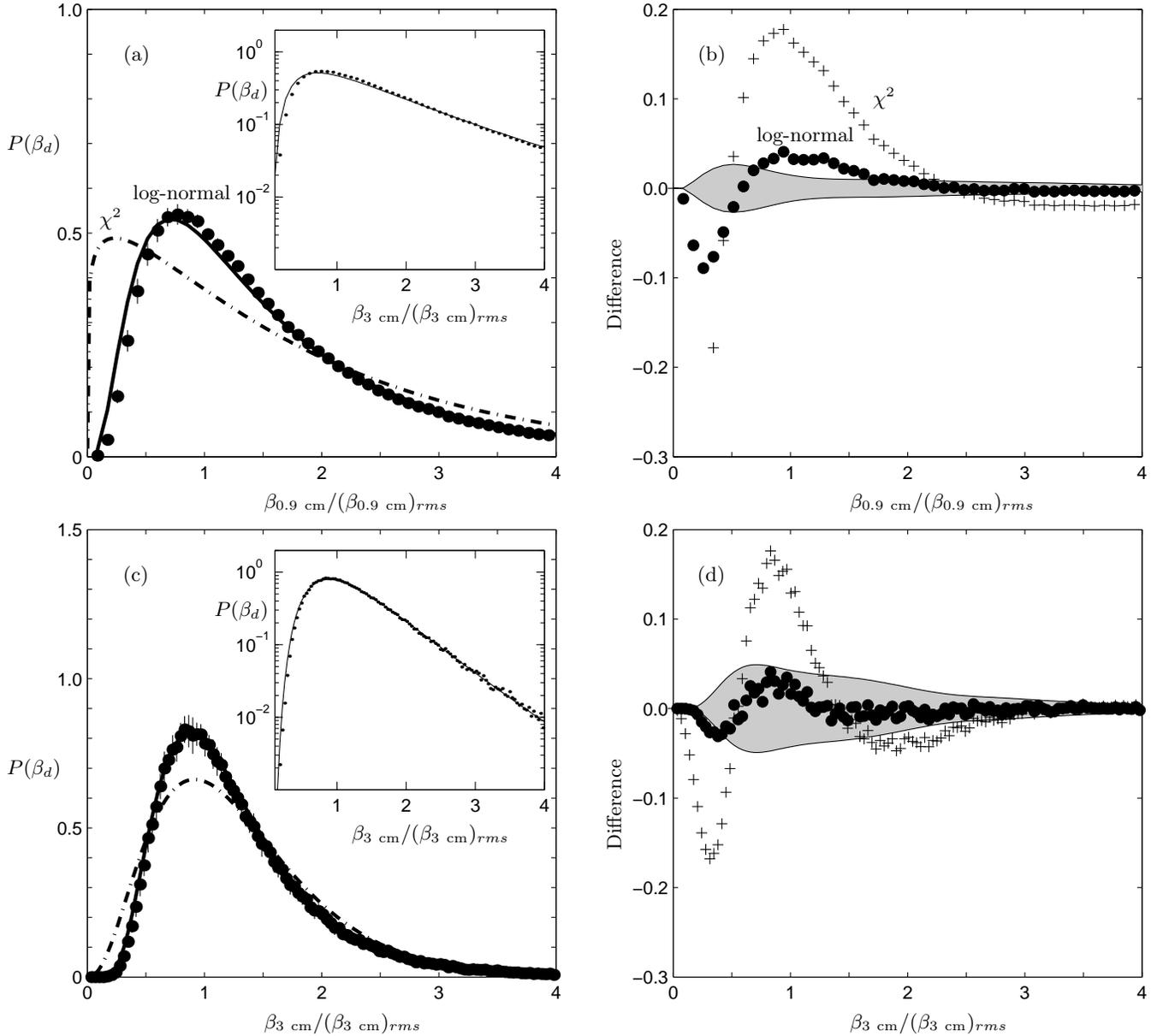}
  \end{center}
\caption{Comparison of ${\chi}^2$ and log-normal distributions to
the experimental distribution for inverse effective temperature in
subsystems of size (a) $d$=0.9 cm and (c) $d$=3 cm. The dash-dotted
lines represent the ${\chi}^2$ distribution, and the solid line
represents the log-normal distribution; both have the same mean and
variance as the 20 independent experiments (error bars correspond to
one standard deviation). The panels on the right, (b) and (d), show
the difference between the experimental PDF for $\beta_d$ and the
${\chi}^2$ (plus signs) and log-normal (bullets) distributions for
(a) $d$=0.9 cm and (c) $d$=3 cm. The shaded area represents the
experimental uncertainty (standard deviation of 20 experiments).}
\label{fig:P_beta_11}
\end{figure*}

Several distributions for inverse effective temperature $\beta_d$
have been discussed by Beck and Cohen~\cite{beck03}. Here we
consider the log normal and $\chi ^2$ distributions, which are most
applicable to turbulent flow. Due to multiplicative processes in
turbulence, the log-normal distribution is often observed for
positive-definite quantities (such as $\varepsilon_r$)
\cite{castaing90,castaing93,LN90,LN175,LN161,LN146}. A log-normally
distributed $\beta_d$ is given by
\begin{equation} \label{eq:super_log_normal}
    P(\beta_d) = \frac{1}{ s (2\pi)^{1/2} \beta_d} \exp
          \left( - \frac{ ( \log \beta_d - m )^2 }{ 2 s^2 } \right)
\end{equation}
where $s = \sqrt{ \ln (1+\sigma^2_{\beta_d}/\bar{\beta_d}^2) }$
and $m = \log (\bar{\beta_d}^2/\sqrt{\bar{\beta_d}^2
+\sigma^2_{\beta_d} })$ are parameters, and $\bar{\beta_d}$ and
$\sigma_{\beta_d}$ are respectively the mean and standard
deviation of $\beta_d$.

The $\chi ^2$ distribution of $\beta_d$ is given by
\begin{equation}
    P(\beta_d) = \frac{1}{\beta_d \Gamma(c)} \left( \frac{\beta_d}{b} \right)^c
               \exp \left( -\frac{\beta_d}{b} \right)
\end{equation}
where $c = \bar{\beta_d}^2/\sigma^2_{\beta_d}$ and  $b =
\sigma^2_{\beta_d}/\bar{\beta_d}$ and $\Gamma$ is the gamma
function. The $\chi^2$ distribution has been observed in recent
measurement of wind turbulence \cite{rizzo04}. The statistical
properties of different distributions are discussed in
\cite{beck03}.

The experimental PDF for $\beta_d$ is compared in Fig.
\ref{fig:P_beta_11} with a log-normal distribution and with a
${\chi}^2$ distribution for two subsystem sizes $d$, 0.9 cm and 3
cm.  The mean $\bar{\beta_d}$ and variance $\sigma^2_{\beta_d}$ of
the inverse effective temperature determine the parameters $s, m, b$
and $c$. For small $d$, the log-normal and ${\chi}^2$ differ
significantly, but for large $d$ they become closer together [Fig.
\ref{fig:P_beta_11}(c) and (d)].  The decrease in variance of
$\beta_d$ with increasing $d$ is similar to decrease observed in the
variance of $\varepsilon_r$ with increasing $r$ ~\cite{LN96}.

The difference between the PDF of $\beta_d$ from experiment and the
${\chi}^2$ and log-normal distributions is shown in Fig.
\ref{fig:P_beta_11}(b) and (d). For $d$=0.9 cm, the log-normal
distribution fits the data within the experimental uncertainty
except small $\beta_d$ regions, while the $\chi ^2$ distribution
deviates from the observations by an amount that is large compared
to the uncertainty. For $d$=3 cm, the log-normal distribution fits
the distribution of $\beta_d$ whereas the $\chi^2$ distribution does
not.

\begin{figure}
 \centering 
 \begin{center}
   \psfrag{P1}{$m$}
   \psfrag{P2}{and} \psfrag{P3}{$s^2/2$}
   \psfrag{dcm}{$d$ (cm)}
   \includegraphics[width=\linewidth]
     {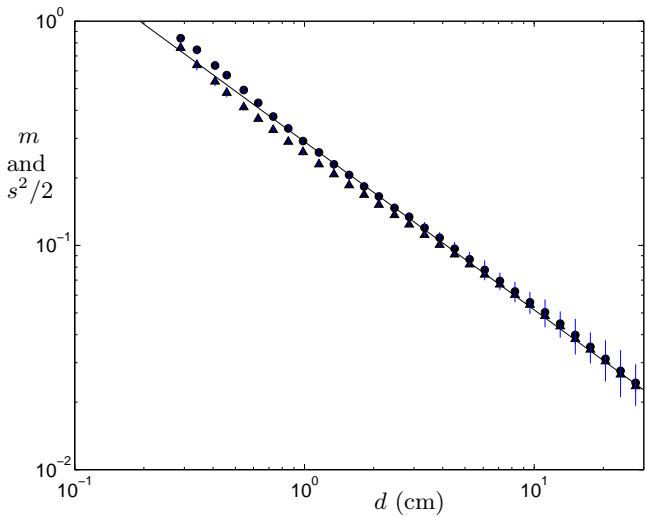}
 \end{center}
 \caption{The parameters $\frac{s^2}{2}$ (circles) and $m$
 (triangles), obtained from fits of the inverse effective temperature $\beta_d$
 (deduced from Couette-Taylor turbulence data) to a log-normal
 distribution, as a function of subsystem size $d$ (see Eq.
 (\ref{eq:super_log_normal}). ($s$ and $m$ are the variance and mean
 of logarithmic inverse effective temperature.) The parameters $\frac{s^2}{2}$
 and $m$ are approximately equal (see text) and are described by a
 power law, $m \propto d^{-3/4}$ (solid line).}
 \label{fig:sup_multi}
\end{figure}

The log normal distribution (\ref{eq:super_log_normal}) involves two
parameters, $s$ and $m$, which depend on subsystem size, as shown in
Fig.~\ref{fig:sup_multi}. This figure suggests a relationship
between $s$ and $m$, $m=\frac{s^2}{2}$, which is supported by a
calculation in Castaing et al. (see Section 4.3.1 in
\cite{castaing90}).

\subsection{Conditional probability and the proper subsystem size}\label{sec:sup_predict}

\begin{figure}
 \centering
 \begin{center}
   \psfrag{4moment}{$4^{th}$ Moment} \psfrag{2moment}{$2^{nd}$ Moment}
   \psfrag{3moment}{$3^{rd}$ Moment} \psfrag{moment of dv}{Moment of $\delta v_r$}
   \psfrag{dcm}{$d$ (cm)}
   \includegraphics[width=\linewidth]
     {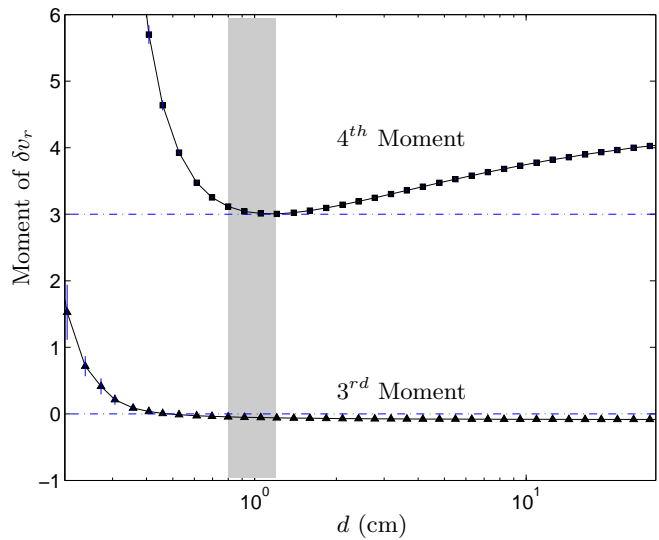}
 \end{center}
 \caption{The dependence of the third and fourth moments of
 $P(\delta v_r)$ on the size $d$ of the subsystems.  For
 sufficiently small $d$, $P(\delta v_r)$ should be Gaussian, which
 means the values of the third and fourth moments should have the
 values zero and three, respectively.  We find that at $d \approx 0.9
\mathrm{cm}$, the conditional distribution
Eq.~(\ref{eq:super_log_normal}) is close to Gaussian (see text). }
 \label{fig:sup_find}
\end{figure}

In the statistical approach of Beck and Cohen, the subsystem size
$d$ should be sufficiently small so that $P(\beta_d)$ is Gaussian,
corresponding to local thermodynamic equilibrium in the subsystems.
However, in practice the $ d \longrightarrow  0 $ limit is
inaccessible because as $d$ becomes very small, the number of data
points becomes too small to allow accurate determination of the
variance of $\beta_d$.  So what is optimal choice of $d$? We address
this question by examining the third moment (skewness) and fourth
moment (kurtosis) of $\delta v_r$, which should be equal
respectively to zero and three for a Gaussian distribution. In
principle we could also examine fifth and higher moments, but
because of the sensitivity of the higher moments to noise, we limit
our considerations to the third and fourth moments. Plotting the
third and fourth moments as a function of $d$, as shown in Fig.
\ref{fig:sup_find}, we find that the optimal value of $d$ for our
data is 1.0-1.2 cm, which is the only range in which the kurtosis is
approximately given by the value for a Gaussian. The skewness is
small and negative for $d>0.5$ cm, but becomes strongly positive for
$d<0.5$ cm, reflecting a cascade of energy to smaller length scales.
We conclude that $d$=0.9 cm is the optimal subsystem size for our
data.

\subsection{Probability distribution of $\delta v_r$} \label{sec:sup_PDF}

\begin{figure}
 \centering 
 \begin{center}
   \psfrag{Pdv}{P($\delta v_r$)} \psfrag{a}{(a)} \psfrag{b}{(b)}
   \psfrag{dvrms}{${\delta v_r}/{(\delta v_r)_{\mathrm{rms}}}$}
   \psfrag{Gaussian}{Gaussian} \psfrag{PthPexp}{$(P_{th} -P_{exp})/P_{th}$}
   \psfrag{3cm}{3 cm} \psfrag{0.3cm}{0.3 cm} \psfrag{0.9cm}{0.9 cm}
   \includegraphics[width=\linewidth]
   {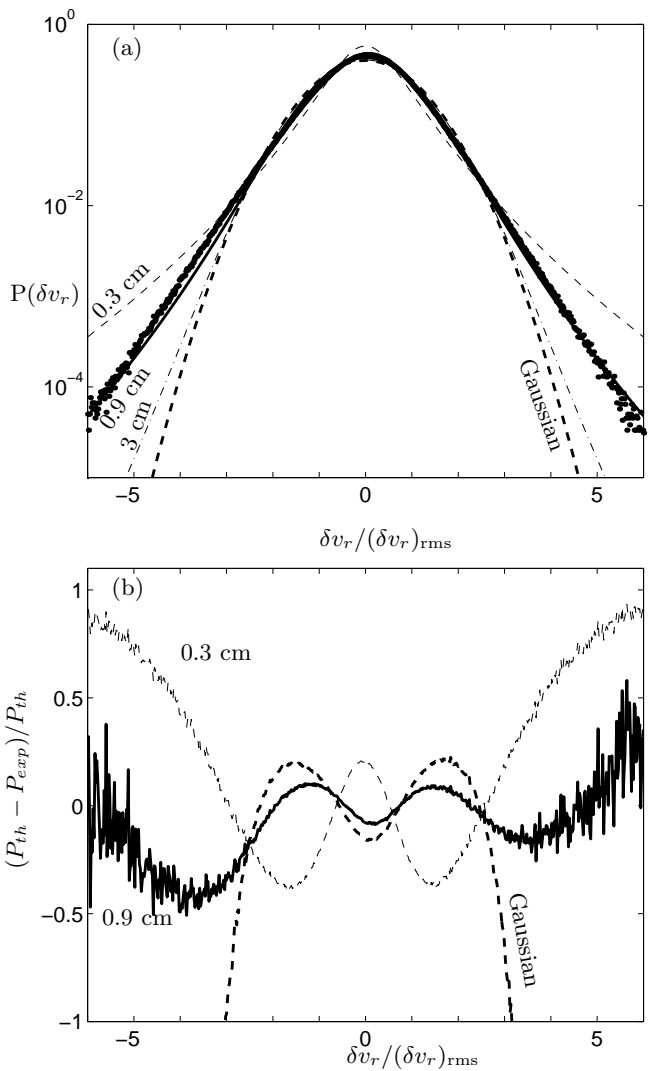}
 \end{center}
\caption{Comparison experimental results (dots) for $P(\delta v_r)$
with the prediction of the Beck-Cohen method for a subsystem with
the optimal size of 0.9 cm (bold line) on semi-log scale in (a) and
relative error between theoretical and experimental values in (b).
For comparison, we also show in (a) and (b) the predictions for
subsystems of size 0.3 cm (thin dashed line) and 3 cm (thin dash-dot
line) and a Gaussian distribution (dashed line).}
 \label{fig:pdf_dv}
\end{figure}

We found a log-normal distribution of $\beta_d$ fits the turbulence
data over a wide range in $d$ (Section \ref{sec:sup_distributions}).
With the log normal distribution of $\beta_d$ for the optimal value
of $d$ (0.9 cm, Fig.~ \ref{fig:sup_find}) and the conditional
Gaussian distribution of ${\delta v_r}$ for that $\beta_d$, we
obtain the probability distribution of ${\delta v_r}$ by the method
of Beck and Cohen,

\begin{eqnarray} \label{eq:sup_p_dv}
P({\delta v_r} ) &=& \frac{1}{2 \pi s} \int_0^{\infty} d\beta_d
\beta_d^{-1/2} \exp \left( - \frac{
( \log \beta_d - m)^2 }{ 2 s^2 } \right) \nonumber \\
                 && \times \exp \left( -\frac{1}{2} \beta_d (\delta
                 v_r)^2 \right),
\end{eqnarray}
where $s$ and $m$ are determined from experiment for the optimal
subsystem size $d$. There is no explicit form for the improper
integral in Eq. (\ref{eq:sup_p_dv}) so we evaluate the integral
numerically, using the limits ($[\min{\beta_d},\max{\beta_d}]$)
measured in experiments instead of the theoretical integral domain,
$[0 , \infty )$.

The results for $P({\delta v_r})$ obtained by numerical integration of
(\ref{eq:sup_p_dv}) are shown in Fig. \ref{fig:pdf_dv}. The data are
described much better by the predicted probability distribution than
by a Gaussian.  The observed approximate power law tails are similar
to the predicted distribution function. 

\subsection{Castaing and Beck-Cohen methods}
\label{sec:sup_comp}

\begin{figure}
 \centering 
\begin{center}
   \psfrag{logbeta}{$\ln \beta_d$}
   \psfrag{loge}{$\ln \varepsilon_r$}
   \includegraphics[width=\linewidth]
   {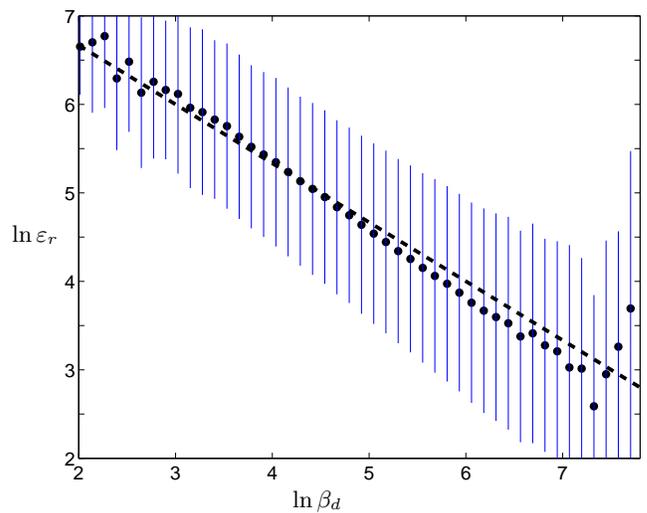}
 \end{center}
\caption{The relation between $\beta_d$ and $\varepsilon_r$. The
solid vertical lines represent standard deviations at a fixed
$\beta_d$ and the dots represent the mean values. The dashed
line is $\beta_d \propto (\varepsilon_r)^{-2/3}$.}
 \label{fig:Functional}
\end{figure}

\begin{figure}
 \centering 
\begin{center}
   \psfrag{lnbeta}{$\ln \beta_d/ (\ln \beta_d)_{rms}$} \psfrag{a}{(a)}
   \psfrag{b}{(b)} \psfrag{Pbeta}{$P(\beta_d | \varepsilon_r)$}
   \includegraphics[width=\linewidth]{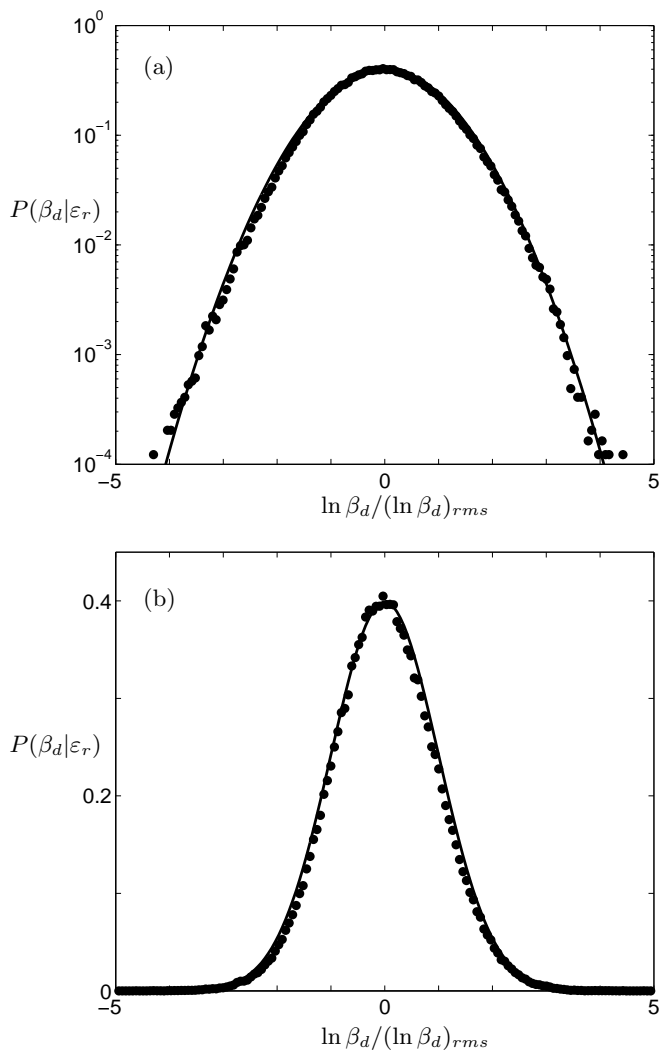}
 \end{center}
\caption{The Gaussian distribution of $\ln \beta_d$ conditioned by
$\varepsilon_r$, plotted on (a) log and (b) linear scales. The solid
lines represent a Gaussian distribution of $\ln \beta_d$, that is,
the log-normal distribution of $\beta_d$. The dots represent the
mean values of $\ln \beta_d$ from experiments.}
 \label{fig:Functional_gauss}
\end{figure}


If the two conditioning quantities in the Castaing and Beck-Cohen
methods ($\varepsilon_r$ and $\beta_d$, respectively) are correlated
as a power-law, through Bayes' theorem the two methods can be seen
to be the same (see Eq. (\ref{eq:sup_uni_2})). With the surrogate
definition of $\varepsilon_r$ as in Eq. (\ref{eq:sup_surro}) and a
proper subsystem size (Section \ref{sec:sup_predict}), we find that
$\beta_d$ and $\varepsilon_r$ exhibit a power-law relation, as Fig.
\ref{fig:Functional} illustrates.  In this sense, the Castaing and
Beck-Cohen methods describe the same PDF of $\delta v_r$ through the
different conditional values which are correlated. Our experimental
observation of a relation $\beta_d \propto (\varepsilon_r)^{-2/3}$
in Fig. \ref{fig:Functional} follows also from a dimensional
analysis,
\begin{eqnarray}
[\beta_d] = \left[ \frac{T^2}{L^2} \right] &=& \left[ L \right]^{-2/3}
\times \left[ \frac{L^2}{T^3} \right]^{-2/3}
\nonumber \\
\Rightarrow \beta_d  &\propto& r^{-2/3}\varepsilon_r^{-2/3} \, ,
\end{eqnarray}
where square brackets $[\cdot]$ denote the dimension of a physical
quantity, $T$ is the dimension of time and $L$ is the dimension of
length.

The probability of $\beta_d$ conditioned to $\varepsilon_r$,
$P(\beta_d | \varepsilon_r) $, is log-normally distributed, as
Fig. \ref{fig:Functional_gauss} illustrates. Our assumption in
Eq. (\ref{eq:sup_uni_3}) holds with the surrogate $\varepsilon_r$ and
$\beta_d$, where $d$ is properly chosen (Section
\ref{sec:sup_predict}). Thus the integral of two log-normal
distributions, $\int P(\beta_d | \varepsilon_r) P(\varepsilon_r) d
\varepsilon_r$, is another log-normal distribution, $P(\beta_d)$.
That is, if $P(\beta_d | \varepsilon_r)$ is a log-normal distribution
with the mean of $\ln \varepsilon_r$, a log-normal distribution of
$\varepsilon_r$ in Castaing's method is equivalent with a log-normal
distribution of $\beta_d$ in Beck-Cohen's method.

\section{Conclusions} \label{sec:sup_conclusion}

Both Castaing and Beck-Cohen methods have been very successful in
describing the non-Gaussian distribution of velocity differences in
turbulence \cite{beck04,castaing90}. Although the relation of
Beck-Cohen's method and Tsallis statistics \footnote {Beck and Cohen
have shown that their method includes Tsallis statistics and other
statistics~\cite{beck03}. A log-normal distribution is
indistinguishable from Tsallis statistics except in long
tails~\cite{beck04}.} to turbulence has been
questioned~\cite{gotoh04,beck04,nauenberg03,tsallis04}, the fit to
data is quite good
\cite{jung04,bodenschatz04a,bodenschatz04b,baroud03b}. We have
presented a method for determining subsystem size in the Beck-Cohen
method, thus eliminating the need for a fitting parameter.

We have also shown that Castaing's method can be converted to
Beck-Cohen method -- the log-normal distribution of $\varepsilon_r$
in Castaing's method gives rise to a log-normal distribution of
$\beta_d$ in Beck-Cohen's method. In that sense, the two methods
describe the non-Gaussian distribution of $\delta v_r$ in the same
way, $P(\delta v_r) = \int$ Gaussian distribution $\times$
log-normal distribution.

The authors thank C. Beck and E.G.D. Cohen for reading the
manuscript and making helpful suggestions, and A.M. Reynolds and E.
Sharon for helpful discussions. This research was supported by the
Office of Naval Research. S. Jung acknowledges the support of a
Donald D. Harrington fellows program.


\bibliography{../../phd/bib_LN.bib,../../phd/pap5.bib,../../phd/bib_K62.bib}

\end{document}